\documentclass[11pt,a4paper]{article}
\usepackage{uereport}
\usepackage[utf8]{inputenc} 
\usepackage{url} 
\usepackage{amsmath,amssymb,amsthm} 
\usepackage{hyperref}
\usepackage{mathtools} 

\usepackage{graphicx}
\graphicspath{{figures/}}

\usepackage{natbib}
\bibpunct{[}{]}{,}{n}{}

\DeclareMathOperator*{\argmax}{arg\,max}

\title{Mitigating Epidemics through \\Mobile Micro-measures}
\author{Mohamed Kafsi, Ehsan Kazemi, Lucas Maystre, Lyudmila
Yartseva,\\Matthias Grossglauser, Patrick Thiran}
\course{School of Computer and Communication Sciences, EPFL}
\address{first.last@epfl.ch}

\begin{document}
\maketitle


\abstract{ 
Epidemics of infectious diseases are among the largest threats to the
quality of life and the economic and social well-being of developing
countries. The arsenal of measures against such epidemics is
well-established, but costly and insufficient to mitigate their impact. In this
paper, we argue that mobile technology adds a powerful weapon to this
arsenal, because (a) mobile devices endow us with the unprecedented
ability to measure and model the detailed behavioral patterns of the
affected population, and (b) they enable the delivery of personalized
behavioral recommendations to individuals in real time. We combine
these two ideas and propose several strategies to generate such
recommendations from mobility patterns. The goal of each strategy is a
large reduction in infections, with a small impact on the normal course
of daily life. We evaluate these strategies over the Orange D4D
dataset and show the benefit of mobile micro-measures, even if only a
fraction of the population participates. These preliminary results
demonstrate the potential of mobile technology to complement other
measures like vaccination and quarantines against disease
epidemics.
%
}

\section{Introduction} 
\label{sec:intro}

Modeling and effectively mitigating the spread of infectious diseases has been a
long-standing public health goal. The stakes are high: throughout human history,
epidemics have had significant death tolls. In the mid-14$^{\text{th}}$ century
\cite{gottfried1985black}, between 30\% and 50\% of Europe's population died due
to the Black Death.  In 1918, the Spanish flu pandemic caused an estimated 50
million deaths worldwide \cite{taubenberger2006influenza}.  More recently, the
2002--2003 SARS pandemic that originated in Hong-Kong and spread worldwide
caused the death of 774 \cite{who2004sars}. These events highlight not only the
scale of the problem but also our vulnerability, past and present.  The
situation worsens in times of crises. A recent example is the ongoing cholera
outbreak in Haiti: it started in 2010, a few months after a major earthquake.
Cholera is a recurring issue in West African countries as well, with many deaths
reported each time. Effective measures against an epidemic require an accurate
and up-to-date assessment of the situation, a very fast response and a strong
coordination, which are colossal organizational efforts under tight time
constraints. To this day, there is no uncontested way of preventing epidemics in
general.  Traditionally, many methods that have been used involve top-down
approaches such as vaccination campaigns, the set-up of medical shelters, travel
restrictions or quarantines \cite{inglesby2006disease}. These methods have
several drawbacks: they are difficult and slow to be put into place, they can be
expensive and also freedom-restrictive. It is clear that any improvement could
have a tremendous impact and translate into significant welfare gains.

In our work, we focus on human-mediated epidemics (transmitted by human contact,
e.g.,\ influenza). For these epidemics, human mobility clearly plays a crucial
role in that it enables the epidemic to travel and spread geographically. We
will explore new mitigation methods and expand the solution space. In
particular, we argue that taking advantage of mobile technology opens up many
possibilities for mitigating the spread of an epidemic in original and
distinctive ways. Importantly, mobile technology is unique in that it is allows
the \emph{personalization} of countermeasures through precise measurements at
the individual level, as well as individualized recommendations. It is this
combination of information extracted from mobile data and subsequent
personalization of prevention advice that opens up novel ways of mitigating an
epidemic.  We envision a mobile service that sends recommendations that invite
the individuals to adapt their behavior, for example by delaying or canceling a
trip. More generally, we formulate subtle, precise and minimally restrictive
personalized behavioral rules that, if followed even partially, will have a
positive global effect on the epidemic.

\subsection{Context and Contributions}

Our work was spurred by the \emph{Data for Development} challenge\footnote{See:
\url{http://www.d4d.orange.com/}. The challenge was launched in mid-2012 and
ended on February 15$^{\text{th}}$, 2013.} organized by France Telecom-Orange, a
global telecommunications operator. Participants in this challenge have access
to data gathered from 2.5 billion calls made by 5 million users in Ivory Coast.
The goal is to find an original and creative use of this data that contributes
towards the social, economic and environmental development of Ivory Coast. Four
different datasets were derived from call detail records (CDRs) recorded over a
period of 5 months, from December 2011 to April 2012.  \citet{blondel2012data}
provide a detailed description of the datasets.  Among these, two are mobility
traces containing the time and the location at which a sample of the users made
their phone calls. In order to protect the users' privacy, the datasets reflect
different trade-offs in terms of the location's accuracy and the time span over
which the trace is provided.  We use this data to build a home location and
time-dependent model of human mobility in Ivory Coast, which allows us to
accurately capture population movements across the country (Section
\ref{sec:mobmod}). These mobility patterns then power the core of our epidemic
model, which allows us to analyze epidemic outbreaks at the level of single
individuals (Section \ref{sec:epidmod}).

Beyond these models, our main contribution is to foster the idea of a mobile
service that sends personal recommendations to help mitigate an epidemic. The
mobile service is an original idea that has several advantages over existing
methods. In particular, we introduce and motivate the concept of
\emph{micro-measures}, individual countermeasures tailored to their recipients'
specific behavior; this new approach is the opposite of the one-size-fits-all
pattern that characterizes most traditional mitigation measures. We present
several concrete such micro-measures and discuss their potential (Section
\ref{sec:mobrec}).  Finally, we empirically evaluate their effectiveness using
our epidemic model and provide some insights into further research directions
(Section \ref{sec:strat}).

\section{Related Work}
\label{sec:relwork}

Infectious diseases, also known as transmissible diseases are one of the the
major causes of deaths in human societies. An epidemic is a rapid and extensive
spread of a transmissible disease that affects many individuals in an area,
community, or population. In order to study epidemics, scientists need to
describe them mathematically, which enables them to predict epidemic outbreaks
and to find strategies for decreasing mortality rates, and hence the costs to
the economy.  In their seminal work, \citet{kermack1927contribution} introduce a
SIR model with three distinct classes of populations: susceptibles, infectives
and recovered. This simple, yet powerful, model is very popular for modeling the
evolution of epidemics in populations.  \citet{hethcote2000mathematics} reviews
different extensions of this model (SIS, SI, SEIS and etc.), as well as
threshold theorems involving measures such as the reproduction number, which is
the average number of secondary infections caused by an infected individual when
in contact with a population of susceptibles.

Instead of modeling an epidemic for the population in a region, it is possible
to increase geographical granularity by dividing the original region in
sub-regions, and then study the SIR model for the population of each region
\cite{arino2003multi, belik2009impact, sattenspiel1995structured,
truscott2012evaluating}.  By assuming that human contacts are responsible for
disease transmission, the disease spread among sub-regions is driven by the
mobility of individuals. \citet{sattenspiel1995structured} take into account the
home region of individuals in order to simulate their mobility.  One of the
simple approaches to modeling population mobility is the gravity model that is
based on two assumptions: Mobility flux between two regions is proportional to
the product of their population's size. It decays as the distance separating
them increases \cite{simini2012universal}. For example, recently,
\citet{rinaldo2012reassessment} study the Haiti cholera outbreak (2010--2011)
and try to predict the next outbreaks of cholera, using the gravity model and
rainfall as drivers of disease transmission. By using a stochastic computational
framework, \citet{colizza2006role} study the epidemic propagation on a larger
scale: They analyse the effect of airline transportation (complete worldwide air
travel infrastructure complemented with census population data) on global
epidemics.

In order to improve the realism of epidemic models, we need to build more
accurate and data-driven mobility models. CDRs collected by cellular services
are used for studying human mobility, because they represent a rich source of
information about mobility \cite{barabasi2005origin, bayir2009discovering,
becker2013human, gonzalez2008understanding, isaacman2011ranges,
tanahashi2012inferring}.  For example, \citet{gonzalez2008understanding} analyze
the trajectory of 100,000 mobile phone users over a six-month period. They find
that human trajectories exhibit very regular patterns, hence we can model each
individual mobility with only a few parameters. \citet{isaacman2012human} model
how a large population move within different metropolitan areas. Because of the
sporadic nature of CDRs, \citet{ficek2012inter} use a Gaussian mixture model to
reproduce probabilistically location of users between two consecutive calls.
Based on the number of unique antennas observed by each user,
\citet{halepovic2005characterizing} assume that some proportion of the
population are static and always stay in their home regions. 

The development of strategies for controlling epidemics such as influenza is one
of the high priorities of global public health policies
\cite{ferguson2006strategies, germann2006mitigation, hufnagel2004forecast,
inglesby2006disease}. SIR models, which incorporate mobility between regions,
represent powerful tools for designing and testing different strategies to
control epidemics.  The quarantine is one of the methods often used to limit the
spread of infectious diseases within human populations. We lack information
however about the effectiveness of quarantine on controlling epidemics.
\citet{sattenspiel2003simulating} use records of the influenza epidemic, which
took place in Canada at 1918-19, to investigate the effect of quarantine. They
show that a quarantine is effective only when mobility is restricted, and that
it depends on its application-time and duration.  In addition to these issues
about the effectiveness of quarantine, there are issues, that include
implementation challenges, economic cost and the violation of civil rights,
especially in the cases of long confinement or isolation from society. Another
way to control epidemics is to vaccinate the susceptible population in a series
of pulses called pulse vaccination \cite{meng2008dynamics, shulgin1998pulse,
stone2000theoretical, zaman2008stability, zaman2009optimal}.  For example,
\citet{zaman2008stability} define a control optimization problem based on the
SIR model. They try to compute the optimum percentage of susceptible population
to be vaccinated at each time. This method requires the vaccination of at least
10 percent of the susceptible population at each time step, in order to make a
small change in the epidemic behaviour of the infectious disease.

\section{Mobile Micro-measures}
\label{sec:mobrec}

Traditional epidemic mitigation methods consist of heavy, top-down approaches
such as blockades, quarantines or large-scale vaccinations. As an alternative,
we suggest that mobile technology could enable a much richer and sophisticated
set of mitigation measures for human-mediated epidemics, which we name
\emph{micro-measures}. Let us illustrate our vision by describing a simple
scenario.

\begin{quotation}
Jean, an 18 year old inhabitant of Ivory Coast living in Northeastern Bouaké,
would like to play pickup football. He knows that a meningitis outbreak just
surfaced in his district, and he does not want to take any risk. Bouaké happens
to be part of a pilot program of a mobile service that helps mitigate the spread
of meningitis. Using his mobile phone, he sends a short request to the service
that instantly computes the following personalized recommendation for him: to
minimize the risk, he should try the football field a few kilometers southwards,
instead of going to the one he is used to. It would be best if he took the
\emph{gbaka} (small bus) in about 17 minutes, this way he would avoid contact
with the kids coming back home from the school nearby.
\end{quotation}

Of course, this scenario presents an idealized and naive view of reality; Jean
might not have a cell phone to begin with, the bus might not have such a precise
schedule, and there might not be alternative locations where people are playing
pickup football. It nevertheless gives an overview of the level of refinement
that can be achieved through personal recommendations. The main properties of
such a service are as follows:

\begin{description}
\item[Personalized.] Recommendations are generated and communicated on an
individual basis. Mobile technology enables this in two ways: first, it allows
for a quantity of valuable behavioral information (such as location and
activity) to be recorded and second, it provides a readily available unicast
communication channel.

\item[Adaptive.] As the epidemic progresses and each individuals' intentions are
discovered, the recommendations are instantly adapted.  The personalization of
mobile micro-recommendations ensures their effectiveness. Such recommendations,
in contrast with most large-scale mitigation efforts,  would typically require
much less time to be set up and would always be in phase with the current state
of the epidemic.

\item[Microscopic.] In contrast with a one-size-fits-all policy that typically
considers an epidemic from a macroscopic perspective, micro-measures tend to
focus on subtle and local changes. These changes, when looked at independently,
are mostly insignificant; but taken together, they result in important global
improvements.

\item[State-independent.]  An additional property of the service is that it is
epidemic-state independent: the recommendation should not depend on whether the
individual is infective or not.  First, it does not require prior knowledge
about the state of an individual: it is often hard to determine precisely when
he becomes infected.  Second, it aligns the incentives: without additional
knowledge, everyone can expect to benefit from following the
recommendation---this might not necessarily be the case when the state is known.
\end{description}


This lays the foundation of our approach but does not yet suggest any concrete
mitigation scheme. Still, there are fundamental questions related to the
feasibility of micro-measures. Under which conditions do small, local changes
(such as an individual agreeing to commute slightly earlier) have a global
impact? How many individuals need to cooperate, and how does this, significantly
alter the dynamics of the epidemic ? An epidemic can often be seen as being
either supercritical (the epidemic grows) or subcritical (it declines).  What
microscopic changes are susceptible to cause a phase transition? Although a
precise characterization of these changes and, by extension, rigorous answers to
these questions are beyond the scope of our work, we intend to show initial
evidence of the relevance of such a mobile service.

\subsection{Concrete Micro-Measures} \label{subsec:micromeas}

Beyond a theoretical argument, our contribution is the description and
evaluation of three concrete strategies we can use to generate micro-measures.
They represent initial baselines for further developments.  Let us first note
that contacts between individuals can broadly be categorized into two groups:
the deliberate contacts are, for example, between family members or at work,
whereas the accidental contacts are formed by random encounters, for instance,
while shopping or commuting.  At a high level, our approach is to maintain
deliberate contacts and rewiring the accidental ones. The idea is to
\emph{weaken} the links in the contact network that form the path through which
the epidemic spreads. By changing its structure, we seek to decelerate the
dynamics and drive the epidemic down to a sub-critical level.

\begin{table}[ht]
\begin{center}
\begin{tabular}{| p{2.8cm} | p{3.7cm} | p{3.6cm} | p{3.7cm} |}
\hline
 & {\sc CutCommunities} & {\sc DecreaseMix} & {\sc GoHome}  \\ \hline \hline 

Knowledge to maintain & List of communities of locations & Social communities of users  &  State of the epidemic across regions \\
\hline
Recommendation & Do not cross community boundaries & Stay with your social circle & Go/stay home \\
\hline
Intuition & Weakening the weak geographical links & Segmenting social communities  & Home is a safe place \\
\hline
\end{tabular}
\end{center}
\caption{We recapitulate the main characteristics of the three strategies we have implemented to mitigate the spread of epidemic.}
\label{tab:emphMeas}
\end{table}

\subsubsection{{\sc CutCommunities} strategy}

\begin{figure}[ht]
  \centering
	\includegraphics[width=0.8\textwidth]{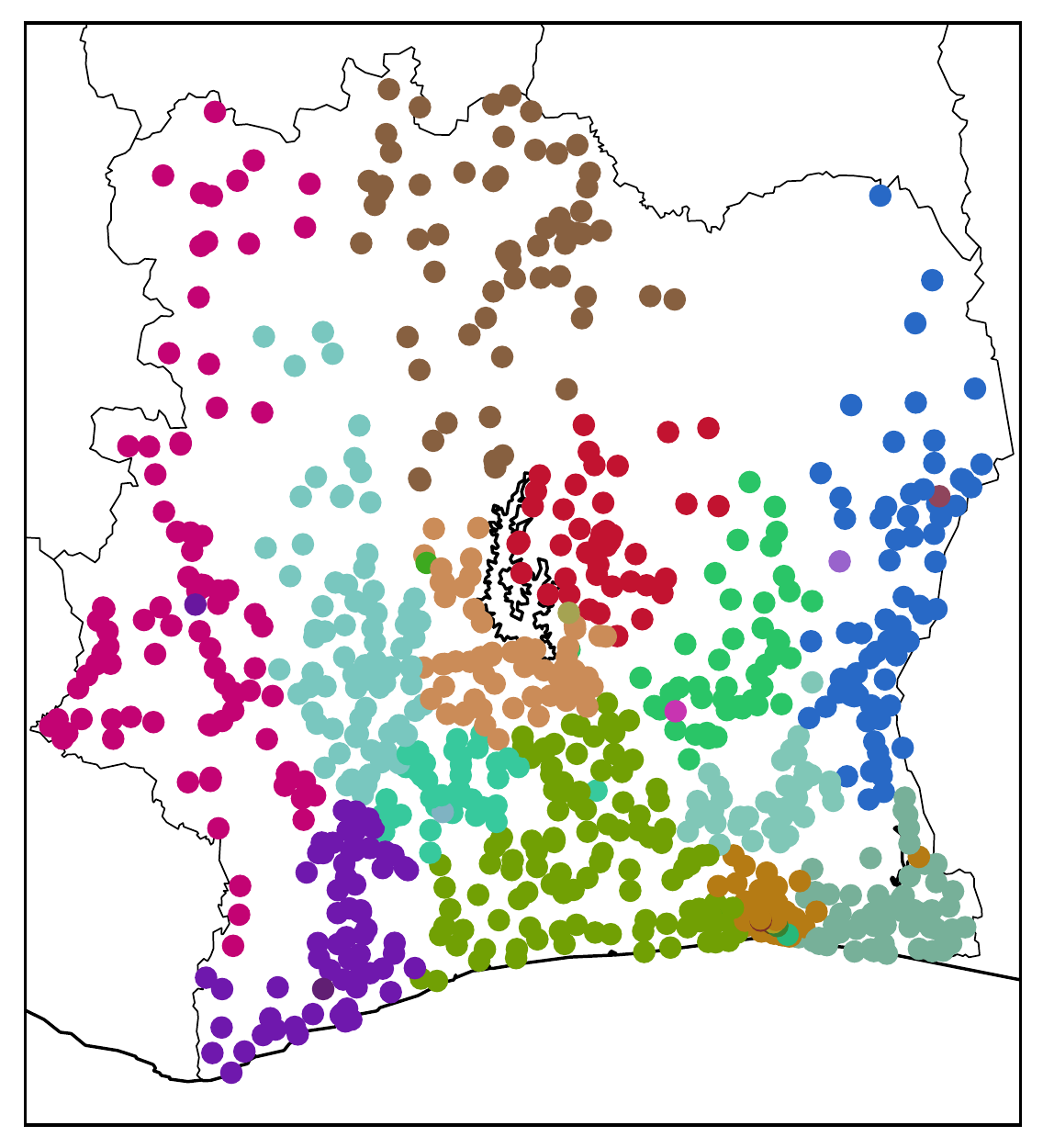}
  \caption{We find 30 communities in the mobility network
  (Section~\ref{sec:strat}) when using the Louvain community detection
  algorithm. It is not surprising that these communities reflects the
  geographical proximity between nodes, as trips are more frequent between close
  antennas than between distant ones.}
	\label{fig:communities}
\end{figure}

It is clear that mobility drives the spread of an epidemic. A straightforward
strategy would therefore be to reduce long-range contacts, be it at the expense
of reinforcing local ones. Uniformly reducing mobility is, however, both
expensive and inflexible. To overcome this, our first strategy, {\sc
CutCommunities}, takes into account \emph{communities of locations} in the
mobility network, and focuses on reducing human mobility over inter-community
links---this is, in a sense, analogous to \emph{weakening the weak links} in the
network. The main difference with a simple blockade is that our strategy is able
to adapt to changes in the network (mobility patterns vary over time, cf.\
Section \ref{sec:mobmod}). In practice, the service operator would maintain a
list of location communities identified through the mobility patterns of its
userbase; when an individual checks whether a trip is safe, the service would
verify whether it crosses community boundaries and, if this is the case,
discourage the individual from making this trip\footnote{As a relaxation of this
counter-measure, we could consider \emph{postponing} the trip instead. Simply by
delaying certain trips, we could prevent harmful interactions between groups of
individuals.  This is analogous to \emph{time-division multiplexing}; a slight
change in the habits of a group of people might significantly change the contact
surface.}. If additional per-location information is available about the current
state of the epidemic, recommendations could be further corrected according to
the strength of the epidemic at the individual's current and projected
locations.

\subsubsection{{\sc DecreaseMix} strategy}

Instead of acting on mobility to segment contacts across location communities,
we also consider segmentation \emph{social communities}. The aim is to separate
individuals \emph{inside} the same location, e.g.,\ by making them visit
different aisles of the same supermarket at different times. Putting in place
such a segmentation is more sophisticated than in the case of mobility, but this
strategy is the perfect example of another extremal point in the solution space.
The service operator would keep a list of social communities and would
communicate a distinctive tag (e.g.,\ a color) to individuals according to their
community. Individuals would access locations differently, depending on their
tag; for example, seating in a theater would be organized in such a way that
contacts between communities are minimized.  We are aware that this strategy
could raise many concerns, because it segregates people, therefore great care
would be needed if it were to be implemented. Despite this, we retain it because
it reflects a different trade-off with respect to {\sc CutCommunities}: instead
of discouraging individuals from going to certain locations where they can be in
contact with everyone, we allow them go everywhere, but restrict the contact
network.

\subsubsection{{\sc GoHome} strategy}

We consider a third case where the service recommends individuals to \emph{go
home}.  The intuition behind this strategy is that we assume that when at home,
the contact rate decreases. Whereas the previous strategies target the
individuals' location or contact network, this one is distinctive in that it
affects the the rate of contact. With information on the progress of the
epidemic across locations, the operator could prioritize sending advice to those
individuals whose cooperation would yield the greatest effect. 
In Section \ref{sec:strat}, we will provide a detailed evaluation of the three
described strategies. Before doing so, we will introduce the mobility and
epidemic models used for our assessments.

\section{Mobility Model}
\label{sec:mobmod}
Because the spread of epidemics depends greatly on the mobility of infected
individuals, and on the locations where they interact with other individuals, a
realistic,  data-driven mobility model is a essential tool for simulating
realistic epidemic propagation.  It should therefore model population mobility,
take into account certain microscopic aspects at the individual level, and still
allow simulations of epidemic propagation to scale up to millions of
individuals. Moreover, it should capture the main differences between the
mobility of different groups of individuals, where a group is constituted of
individuals exhibiting similar mobility profiles.  To construct a mobility model
that fulfils these requirements, our intuition is : The home location of
individuals strongly shapes their mobility patterns because the places they
visit regularly e.g., their workplaces, schools or the shopping centers, depend
on the proximity to their home. Typically, we expect the most visited location
(home) and the second most visited location (school, university or work) to be
geographically close to each other. In addition to this geographical aspect,
mobility is strongly time-dependent: Individuals commute between home and work
during the weekdays, with a substantial change in their travel behavior during
the weekends. 

\begin{table}
\begin{center}
\begin{tabular}{| l | c | l |}
\hline
Definition & Domain & Explanation    \\ \hline \hline 
$\mathcal{A}$ = \{1, \ldots, 1231\} & -  & Set of antennas  \\
$\mathcal{SP}$ = \{1, \ldots, 255\} & -  & Set of sub-prefectures \\
$k$ & $\mathbb{N}$  & Time resolution \\ 
\hline
$sp_{\text{home}}(u)$ & $\mathcal{SP}$  & Home sub-prefecture \\ 
$a_{\text{home}}(u)$ & $\mathcal{A}$  & Home antenna \\ 
$X(n)$ & $\mathcal{A}$  & Antenna \\ 
$t(n)$ & $\mathbb{N}$ & Absolute time  \\   
$h^k(n)$ & $\{1,\ldots,k\}$ & Period of the day   \\ 
$d(n)$ = day($t(n)$) & $\{1,\ldots,7\}$ & Day of the week \\  
$w(n)$ = weekday($t(n)$) & $\{0, 1\}$ & Day type: weekday or weekend \\
\hline
\end{tabular}
\end{center}
\caption{List of the definition and domain of the variables relative to user
$u$, as well as those describing his $n^{\text{th}}$ visit.}
\label{tab:notation}
\end{table}

Building on this, we make the assumption that the individuals that share the
same home-location exhibit a similar mobility pattern. Therefore, we construct a
location and time-based mobility model that depends on the variables presented
in Table~\ref{tab:notation}. The conditional distribution of the location $X(n)$
of user $u$ depends on his home antenna $a_{\text{home}}(u)$, but also on the
time of the visits $(h^k(n),w(n))$:
\begin{equation}
\label{eq:home_antenna_time_dist}
p\left( X(n)|u, t(n)\right) = p\left( X(n)|h^k(n),w(n),a_{\text{home}}(u)\right).
\end{equation}
First, we choose the the time resolution $k=3$ in order to divide the day in 3
distinct periods: Morning (6 am to 1 pm), afternoon (1 pm to 8 pm) and night (8
pm to 6 am). Second, conditioning on the parameter $w(n)$ allows us to
distinguish between weekdays and weekends. Finally, the home antenna
$a_{\text{home}}(u)$ of user $u$ is defined as the most visited antenna during
the night period.  Consequently, given the period of the day, the day type and
the home antenna of user $u$, the distribution of the location that he might
visit~\eqref{eq:home_antenna_time_dist} is a multinomial distribution with
$|\mathcal{A}|$ categories. 

\subsection{Learning and Evaluating Mobility Models}
In order to build our model from data, we analyse {\tt SET2}, one of the
datatsets provided by France Telecom-Orange ~\cite{blondel2012data}. It contains
high-resolution trajectories of 500,000 users, observed over a two-week period.
We focus on this datatset, as it offers the highest geographical resolution :
Individuals' locations are observed across antennas. To avoid having to deal
with users whose location samples are very sparse, we consider only the users
who visited more than 1 antenna and made on average more than 1 call per day.
In order to evaluate the realism of our mobility model, we separate the data
into two parts: For each user, we put $90\%$ of the calls in the training set
and the remaining $10\%$ in the test set. First, we build a mobility model by
learning from the training set by using a maximum likelihood estimator .  Then,
we evaluate our mobility model by computing the average log-likelihood of the
calls belonging to the unseen test set. The average log-likelihood reflects how
well our model generalizes to unseen data. As the test set might contain some
locations not visited by a given class of users in the training set, the maximum
likelihood estimate of the distribution~\eqref{eq:home_antenna_time_dist}
assigns zero probability to these observations. We cope with this by assuming
that the distribution~\eqref{eq:home_antenna_time_dist} is a multinomial
distribution drawn from an exchangeable Dirichlet distribution, which implies
that the inferred distribution~\eqref{eq:home_antenna_time_dist} is a random
variable drawn from a posterior distribution conditioned on the training data. A
more detailed description of this smoothing procedure is given by
\citet{blei2003latent}.

We tested several variants of mobility models by varying their structure and
parameters (time resolution, day of the week, etc). To have three representative
baseline models for comparison, we choose three predictors out of the several
variants we tested.

The first baseline model is a time-based mobility (TM) model 
\begin{equation}
p\left( X(n)|u, t(n)\right) = p\left( X(n)|h^k(n),w(n)\right),
\end{equation}
where all mobile-phone users exhibit the same time-dependant geographical distribution.  
The second baseline is a location-dependent first order Markov chain (MC)
\begin{equation}
p\left( X(n)|u,t(n),X(n-1),\ldots,X(0)\right) = p\left( X(n)|X(n-1)\right),
\end{equation}
where the current location of a user depends only the location he visited just before. 
The third baseline is a time and sub-prefecture dependant mobility model (SPM)
\begin{equation}
p\left( X(n)|u, t(n)\right) = p\left( X(n)|h^k(n),w(n),{sp}_{\text{home}}(u)\right),
\end{equation}

where the home of a user is represented by a sub-prefecture instead of an
antenna. This implies a more important aggregation of users, where two users who
share the same home sub-prefecture, have the same mobility pattern. 

\begin{table}
\begin{center}
    \begin{tabular}{|c|c|}
        \hline
        Mobility model & Average log-likelihood \\ \hline
        Our model & -1.07 \\ 
        SPM & -1.67 \\
		TM & -2.9 \\ 
		MC & -6.49 \\ \hline
    \end{tabular}
\end{center}
\caption{Log-likelihood of the unseen data from the test set. Our mobility model
significantly outperforms the baseline models since its predictive power, with
respect to the test set, is higher.}
\label{tab:llikelihood}
\end{table}

The experimental results are shown in Table~\ref{tab:llikelihood}. The first
order Markov chain (MC) performs the worst. This is not surprising since the
time difference between two call records varies greatly, ranging from a few
minutes to a few days. The location associated with a call made in the past few
hours or days does not necessarily affect the current location. As the location
data is sporadic, it is not surprising than any model that learns from
transitions performs poorly, and is outperformed by time-based models.  Our
model performs the best; and by comparing it to the time-based model (TM), we
realise that knowing the home-locations of users enhances the predictive power
of the mobility model. Moreover, the granularity of home locations is crucial:
Our model significantly outperforms the sub-prefecture dependent mobility model
because it has a finer granularity of the home-location.

A realistic mobility model is an essential building block of a realistic
epidemic propagation model because mobility drives population flows between
regions, and therefore the geographical proximity between individuals. In the
next section, we introduce the epidemic model we use to simulate a local
epidemic propagation. 

\section{Epidemic Model}
\label{sec:epidmod}
Building up on the mobility, this section introduces our epidemic model.  It is
based on a discretized, stochastic version of the \emph{SIR} model
\cite{kermack1927contribution}; Tables~\ref{tbl:epidparams} and
\ref{tbl:epidnot} provide an overview of the different parameters and quantities
used throughout the section. We assume that the size of the population ($N$
individuals) remains constant---there are no births nor deaths, a reasonable
assumption if the time horizon is limited to at most a few months. Under the
\emph{SIR} model, an individual can be either \emph{susceptible} to the disease,
\emph{infective}, or \emph{recovered} from the disease and immunized against
further infections\footnote{In the literature, this state is sometimes known as
\emph{removed}. The important point is that they do not participate in the
epidemic anymore.}. We assume that most of the population is initially
susceptible, except for a small number of infective individuals that form the
seed of the epidemic. Individuals successively go through the susceptible,
infective and recovered states; a desirable outcome would have many individuals
stay susceptible without ever becoming infective.  The basic \emph{SIR} model
assumes \emph{random mixing} of the whole population: any individual meets any
other one with a uniform probability.  In our model, we relax this strong
assumption by taking into account the \emph{mobility}.  We spread the population
across $M$ regions; each region bears its own \emph{SIR} process where the
corresponding meta-population mixes at random.  These regional processes are
independent and isolated, and the only way the epidemic crosses regional
boundaries is through human mobility \cite{keeling2010individual}. In summary,
regional interactions take place uniformly at random, whereas global
interactions are shaped by the individuals' mobility.

\begin{table}[ht]
  \centering
  \begin{tabular}{l|p{10cm}}
    $N$   & total population \\
    $M$   & number of regions \\
    $N^*_i$ & initial population of region $i$, where $i \in \{1, \ldots, M\}$ \\
    $L$   & number of different mobility classes \\
    $\beta$ & contact probability \\
    $g$   & recovery probability
  \end{tabular}
  \caption{Parameters of the epidemic model.}
  \label{tbl:epidparams}
\end{table}

\begin{table}[ht]
  \centering
  \begin{tabular}{l|p{10cm}}
    $c_l$ & mobility class $l$, where $l \in \{1, \ldots, L\}$ \\
    $\mathbf{S}_i$ & distribution of the number of susceptible
    individuals in region $i$ across classes. $\mathbf{S}_i = (S_{i, c_1},
    \ldots, S_{i, c_L})$ \\
    $\mathbf{I}_i$ & distribution of the number of infected individuals
    in region $i$ across classes. $\mathbf{I}_i = (I_{i, c_1}, \ldots, I_{i,
    c_L})$ \\
    $\mathbf{R}_i$ & distribution of the number of recovered individuals
    in region $i$ across classes. $\mathbf{R}_i = (R_{i, c_1}, \ldots, R_{i,
    c_L})$ \\
    $S_i$ & number of susceptible individuals in region $i$, equal to
    $\|\mathbf{S}_i\|_1$ \\
    $I_i$ & number of infected individuals in region $i$, equal to
    $\|\mathbf{I}_i\|_1$ \\
    $R_i$ & number of recovered individuals in region $i$, equal to
    $\|\mathbf{R}_i\|_1$ \\
    $N_i$ & population of region $i$, where $i \in \{1, \ldots, M\}$ \\
    $\lambda_i$ & infection probability for region $i$. $\lambda_i = \beta \frac{I_i}{N_i}$ \\
  \end{tabular}
  \caption{Notation for various quantities related to the epidemic.}
  \label{tbl:epidnot}
\end{table}

\subsection{Local Epidemic Dynamics}

In order to work at the individual level, we adapt the classic deterministic
\emph{SIR} model in order to have a discrete-time stochastic variant. The
contact probability $\beta$ and recovery probability $g$ are constant across all
regions\footnote{These quantities are \emph{rates} in the continuous time
\emph{SIR} model. In order to carry over the characteristics of the \emph{SIR}
model to our discretized version, we need to ensure that the sampling interval
is short enough to ensure that $\beta, g < 1$.}. For a region $i \in \{1,
\ldots, M\}$ we compute, at each time step, the force of infection $\lambda_i =
\beta \tfrac{I_i}{N_i}$. This quantity represents the probability of making a
contact that results in an infection. During a time step, every susceptible
individual gets infected independently at random with probability $\lambda_i$,
while every infective individual recovers independently at random with
probability $g$. If we denote by $\Delta X_i$ the variation of $X_i, X_i \in
\{S, I, R\}$ after one time step, it is easy to see that
\begin{align*}
\mathbb{E}(\Delta S_i) &= -\lambda_i S_i \\
\mathbb{E}(\Delta I_i) &= \lambda_i S_i - g I_i \\
\mathbb{E}(\Delta R_i) &= g I_i,
\end{align*}
which are the expected difference equations for the \emph{SIR} model under the
random mixing assumption. We note that our model has many similarities with that
of \citet{colizza2007predictability}, used to model the SARS pandemic.

\subsection{Implementation}

\begin{figure}[ht]
  \centering
	\includegraphics[width=0.45\textwidth]{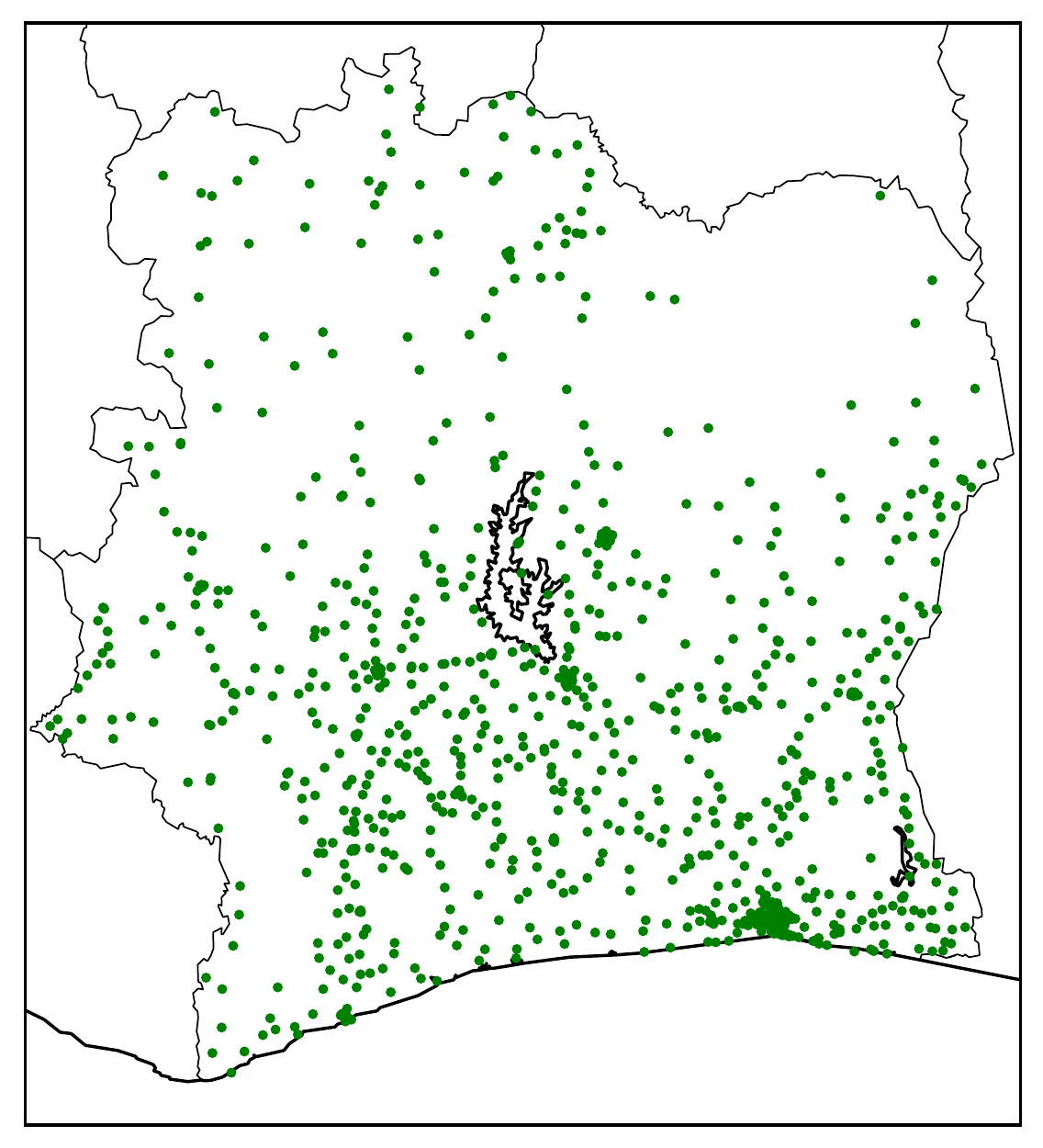}
	\includegraphics[width=0.45\textwidth]{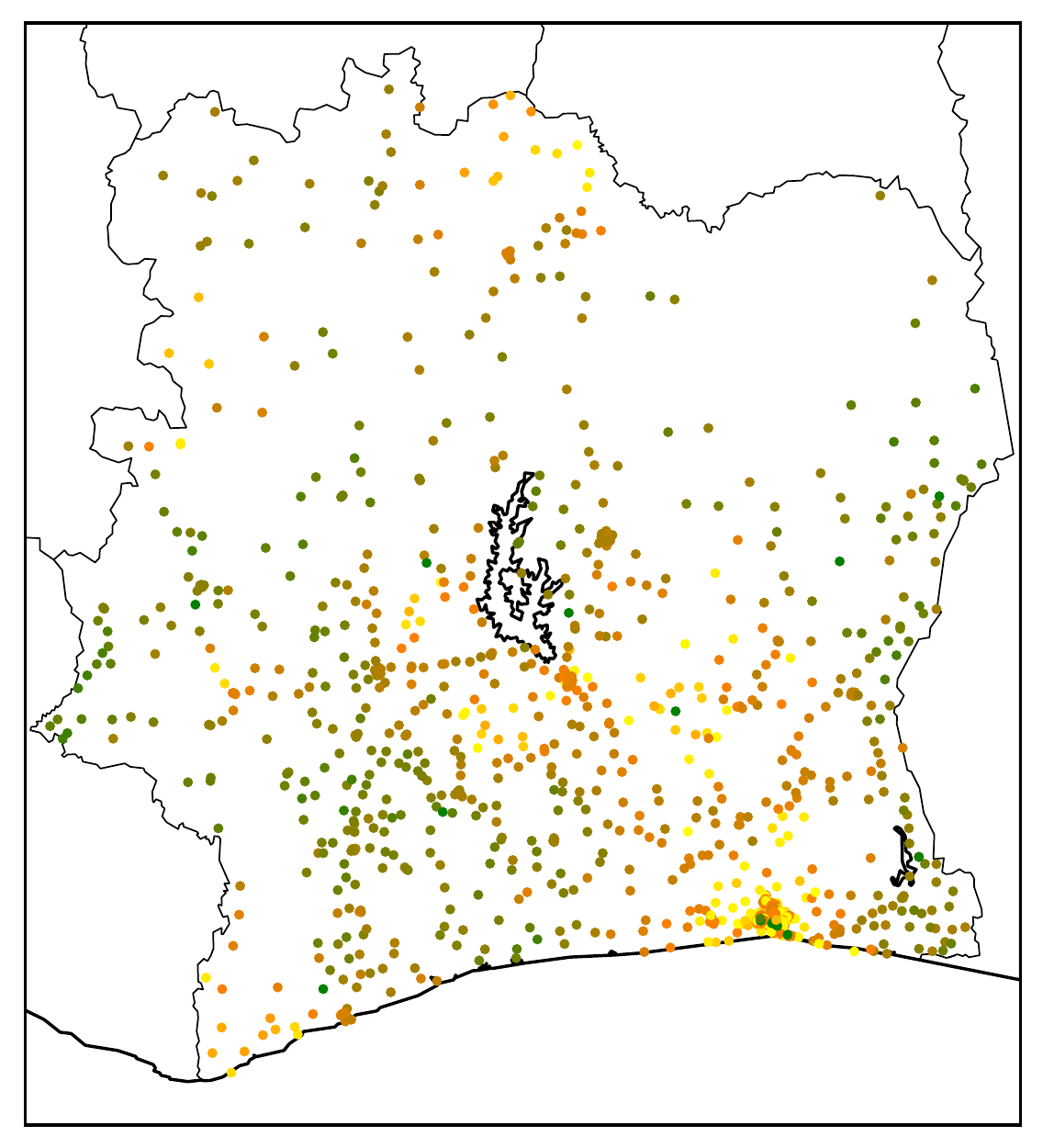}
  \caption{Snapshots of a sample epidemic process where each dot represents a
  region (here, the surroundings of an antenna). Colors indicate the relative
 proportion of infective individuals. Initially, just a few
  individuals form a seed of infectives (left). A little more that 9 days later,
  the epidemic has spread over most of the country (right).}
	\label{fig:epidspread}
\end{figure}

To allow for distinctive mobility patterns across the population, individuals
belong to one out of $L$ classes $\{c_1, \ldots, c_L\}$ that fully characterize
their mobility patterns. In accordance with the mobility model (Section
\ref{sec:mobmod}), the individuals' class is determined by their home antenna.
The implementation is best understood when decomposed into two distinct,
successive phases: a \emph{mobility} phase where individuals can move between
regions, and an \emph{epidemic} phase where individuals get infected or recover.

\begin{description}
\item[Mobility phase] We consider every individual. Suppose the individual is in
region $i$; the mobility model assigns a new region $j$ according to its
mobility class.  If $i \ne j$ we update the vectors $\mathbf{X}_i$ and
$\mathbf{X}_j$ accordingly, where $\mathbf{X} \in \{\mathbf{S}, \mathbf{I},
\mathbf{R}\}$ depends on the current state of the individual.

\item[Epidemic phase] We consider every region $i \in \{1, \ldots, M\}$. We
begin this phase by updating the infection rate $\lambda_i$ given the current
values of $N_i$ and $I_i$. Every infected individual then recovers with
probability $g$, while every susceptible individual gets infected with
probability $\lambda_i$.  $\mathbf{S}_i$, $\mathbf{I}_i$ and $\mathbf{R}_i$ are
updated accordingly.
\end{description}

This process is repeated until the end of the period of interest.

\section{Empirical Evaluation}
\label{sec:strat}

Next, we use our models to test the strategies previously described in Section
\ref{sec:mobrec}. Before evaluating our strategies, we first explain how the
epidemic model is parameterized and how epidemic spreads are quantitatively
characterized.

\subsection{Model Parameters and Evaluation Metrics}

In order to be consistent with our mobility model, the epidemic model defines
regions to be the area surrounding the antennas ($M = 1231$). Hence, we will use
the words \emph{region} and \emph{antenna} interchangeably. As an individual's
mobility is tied to his home antenna, we distinguish among $L = 1231$ different
classes.  To initialize the population attached to each antenna, we use data
from the AfriPop project \cite{tatem2010afripop} which provides us with Ivory
Coast population figures at the hectare level; to account for the fact that not
every individual is mobile, we allow only 55\% of the population to move during
the mobility phase\footnote{This distinction is rather crude and could certainly
be further refined. However we deemed it to be sufficient for our purposes.},
which roughly corresponds to the proportion of the population in the 15-to-64
age bracket \cite{un2010population}. Days are divided into three time steps in
order to match the mobility model\footnote{Notice that this is not a formal
requirement.  We use this subdivision mainly for simplicity.}, and the typical
time horizon is between 100 and 400 time steps (i.e.\ 1--4 months). Contact and
recovery probability are usually set to $\beta = 1$, respectively $g = 0.5$;
Although these synthetic values do not directly match any well-known disease,
they are still qualitatively close to realistic cases, such as influenza. All
our simulations start with a seed set of 23 infectives distributed across 5
antennas\footnote{In the datasets provided by France Telecom-Orange, these
antennas have the following identifiers: 57, 146, 330, 836, 926.} in the
Attécoubé district of Abidjan.

\begin{figure}[ht]
  \centering
	\includegraphics[width=0.8\textwidth]{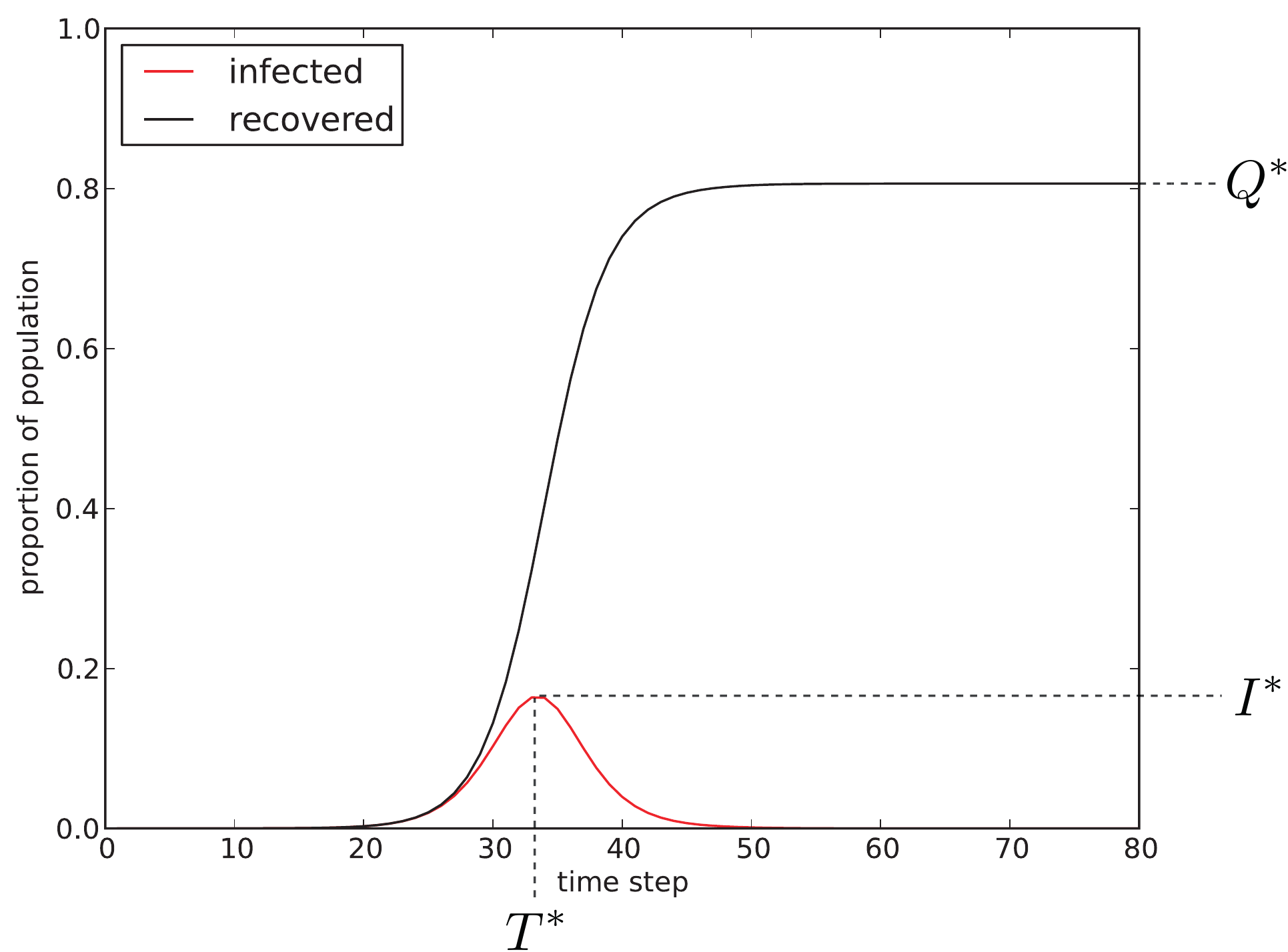}
  \caption{Metrics used to evaluate the effectiveness of mitigation strategies.
  $I^*$ indicates the magnitude of the epidemic's peak, $T^*$ the time at which
  the peak happens, and $Q^*$ describes the asymptotic number of individuals
  that got infected and recovered.}
	\label{fig:metrics}
\end{figure}

In order to quantify the difference between epidemic spreads, we propose three
metrics to evaluate the effectiveness of our mitigation strategies.
Figure~\ref{fig:metrics} shows how these quantities are related to the
epidemic's evolution over time.  For notational clarity, let $X = \sum_{i = 1}^M
X_i, X \in \{S, I, R\}$ be the total number of individuals in each state over
the country as a whole. As these quantities evolve over time, they are functions
of the time step $n$. The first metric is the \emph{size of the largest
outbreak} or, equivalently, the maximal proportion of infective individuals,
$$ I^*= \max_{n} \frac{I(n)}{N} .$$
The reasoning behind this metric is self-evident: in most cases, the larger the
proportion of infective individuals, the more difficult the control of the
epidemic. It is also, broadly speaking, a good indicator of the epidemic's
strength. Our second metric is closely related to the first one, but considers
the complementary dimension: it measures the \emph{time of the largest
outbreak},
$$ T^*= \argmax_{n} I(n).$$
Delaying the moment at which the epidemic reaches its peak allows individuals
and governments to have enough time to adapt their behavior, respectively, to
deploy measures.  Finally, our last metric captures the tail behavior of the
epidemic: it measures the \emph{final proportion of recovered users},
$$ Q^* = \lim_{n \to \infty} \frac{R(n)}{N}.$$
Note that we would like to minimize this metric. After the epidemic dies out,
all individuals are either recovered or susceptible, and a low proportion of
recovered individuals means that a high percentage of the population did not go
through the infective state at all.

\subsection{Results}

We now take a closer look at our three proposed strategies. We
will describe how we instantiate them and we provide qualitative and
quantitative assessments with respect to their effectiveness.

%

\subsubsection{{\sc CutCommunities} strategy}

\begin{table}
  \begin{center}
  \begin{tabular}{| l | c | l |}
  \hline
  $p$ & Affected movements & Maximum   \\ \hline \hline 
  $0.90$ & $10.91 \%$ & $21.38 \%$  ($ts = 42$) \\
  $0.99$ & $12.57 \%$ & $22.91 \%$  ($ts = 51$) \\
  $1.00$ & $5.32  \%$ & $12.20 \%$  ($ts = 33$) \\
  \hline
  \end{tabular}
  \end{center}
  \caption{Proportion of movements affected when using the {\sc CutCommunities}
  strategy for three different values of the compliance probability $p$. We
  indicate the overall average over the $80$ time steps, as well as the maximum
  value.}
  \label{tab:cutcommunities}
\end{table}

\begin{figure}[ht]
  \centering
	\includegraphics[width=0.49\textwidth]{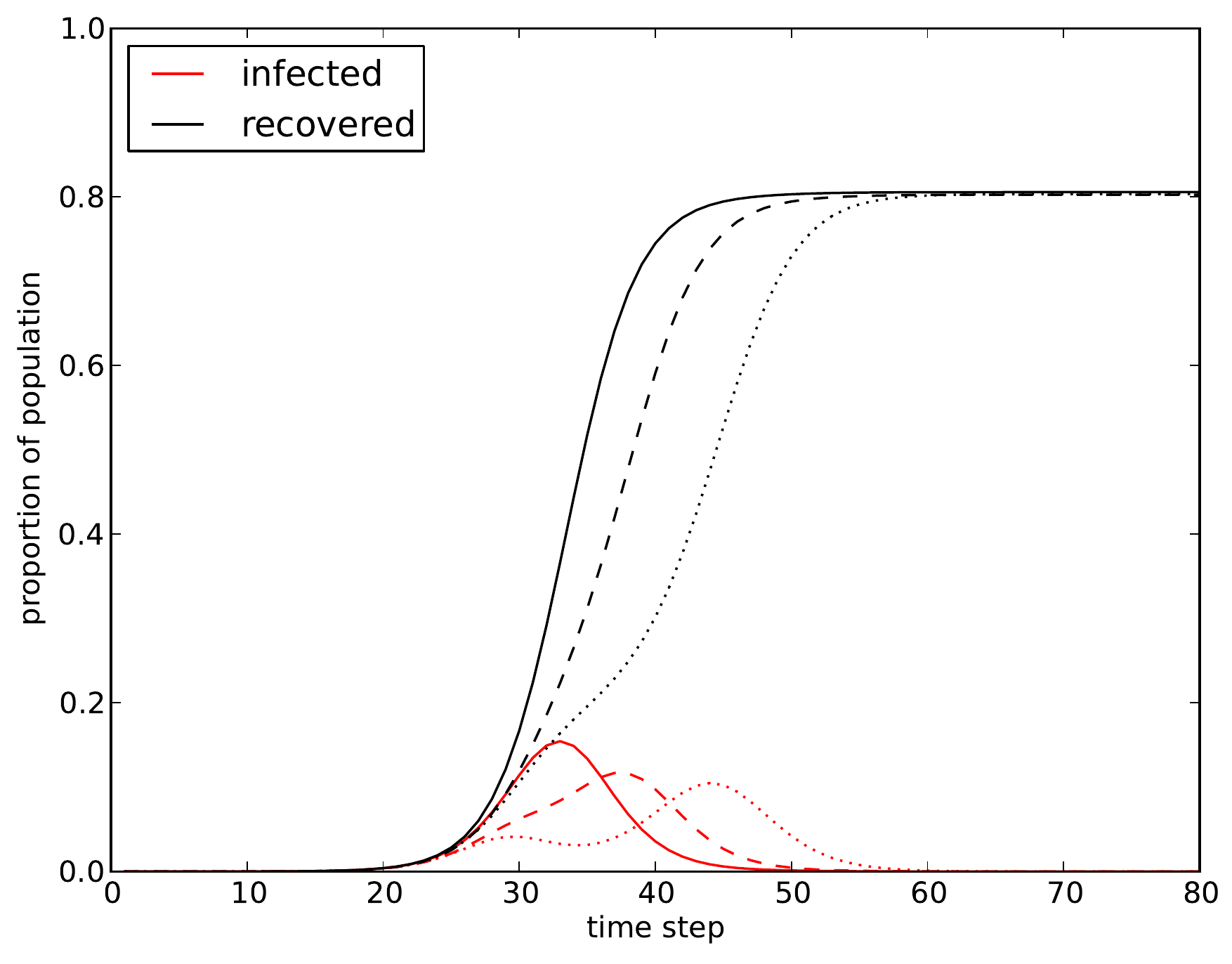}
	\includegraphics[width=0.49\textwidth]{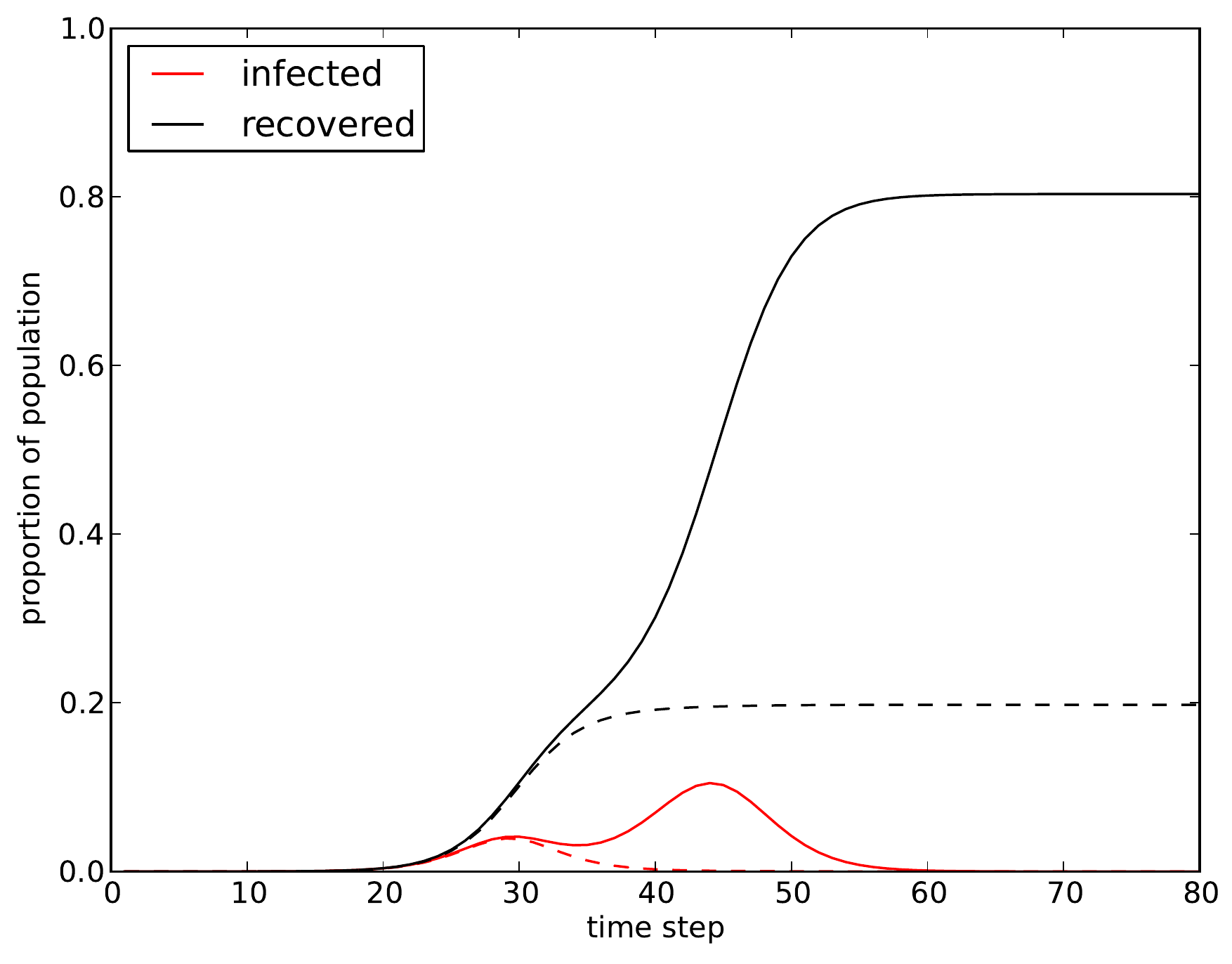}
  \caption{Shape of the epidemic under the {\sc CutCommunities} strategy, $\beta
  = 1.0$, $g = 0.5$. On the left: solid lines represent the baseline ($p = 0$),
  dashed lines $p = 0.9$, dotted lines $p = 0.99$. On the right, we compare $p =
  0.99$ (solid) to a complete blockade ($p = 1$, dashed).}
	\label{fig:cutcommunities}
\end{figure}

The first strategy divides the country into location communities according to
the network of mobility. We consider the weighted, undirected graph where nodes
represent antennas, and edge weight is equal to the average number of trips
between the two endpoints (regardless of direction). We use the Louvain
community detection algorithm \cite{blondel2008fast}; Figure
\ref{fig:communities} shows the 30 identified communities. It is
interesting---but not surprising---to note that the communities are roughly
geographicaly based\footnote{As a sidenote, we ran the Louvain method on a
number of other graphs generated from the datasets provided for the D4D
challenge, including one derived from {\tt SET1} representing total
antenna-to-antenna communications. The communities always displayed the same
geographical clustering. Furthermore, we observed that mobility communities seem
to be correlated to phone call communities.}. This confirms our hypothesis
stating that there are \emph{geographical weak links}. Micro-measures are then
generated as follows: when an individual checks whether a trip is safe, the
service first verifies whether the trip crosses community boundaries and whether
the current or projected locations are affected by the disease; if both of these
conditions are met, the individual is discouraged from making the trip. The
recipient then complies with probability $p$.

Figure~\ref{fig:cutcommunities} shows the effect of {\sc CutCommunities} for
different values of $p$. Compared to the baseline ($p = 0$), the strategy
affects the size $I^*$ and the time $T^*$ of the epidemic's peak. However, it
does not change much the tail behavior: $Q^*$ stays constant at around $0.8$,
except for the degenerate case where $p = 1$, which represents a blockade around
the community initially infected.  We also observe that there seem to be two
infection phases, made progressively more apparent as $p \to 1$, and that the
blockade removes the second phase; these two phases correspond to infections
happening inside, respectively, outside the initially infected community. Recall
that this strategy only sends micro-measures to a fraction of the individuals,
those who cross community boundaries---a case that by definition should not
happen too often. It is therefore interesting to consider the number of trips
actually \emph{canceled} as a result: Table~\ref{tab:cutcommunities} lists the
average and maximal proportion for different values of $p$. The numbers are
quite low\footnote{That these proportions are lowest when $p = 1$ is due to the
fact that the epidemic is local to the infective seeds' community}, suggesting
that the communities form a natural partitioning of the regions. In conclusion,
this strategy does not affect the asymptotic behavior of the epidemic but
significantly shifts its peak.  Altogether, it justifies the relevance of
mobility-based geographical communities as a data source to generate
micro-measures.

\subsubsection{{\sc DecreaseMix} strategy}

\begin{figure}[ht]
  \centering
	\includegraphics[width=0.8\textwidth]{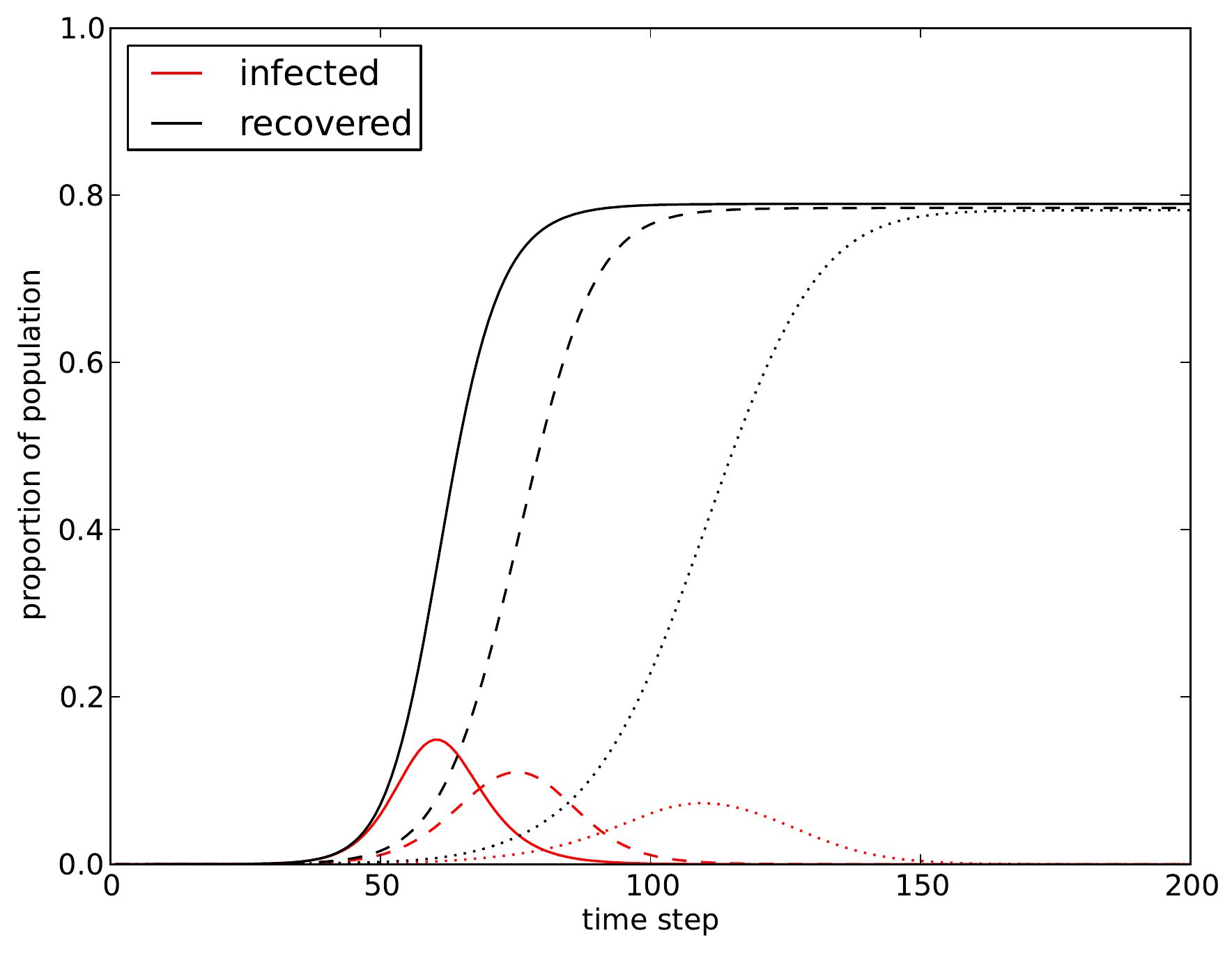}
  \caption{Shape of the epidemic under the {\sc DecreaseMix} strategy averaged
  over 10 runs, $\beta = 1.0$, $g = 0.5$, for different values of the mixing
  parameter. Solid lines correspond to $q = 1.0$, dashed ones to $q = 0.1$,
  dotted ones to $q = 0.01$.}
	\label{fig:decmixres}
\end{figure}

Recall that this strategy assigns tags to individuals according to the social
community to which they belong and segregates contacts across social
communities.  A service operator might use the call graph (i.e.\ the social
network derived from who calls whom) to infer social communities in the
population; unfortunately, we do have access to such data\footnote{The data
provided for the Orange D4D challenge does include a dataset consisting of
myopic views of the call graph. {\tt SET4} is a sample of \emph{egonets}, i.e.\
balls of radius two centred at a particular user. However, this dataset did not
yield anything useful for our purposes.}. In order to quantify the effectiveness
this strategy, we proceed as follows. Similarly to our mobility model, we make
the assumption that the individual's community $C$ is determined by his home
antenna.  The {\sc DecreaseMix}  strategy do not decrease the total number of
contacts; instead it rewires contacts across communities to contacts inside the
community.  This is done by splitting the contact probability to into
intra-community and inter-community contact probabilities and introducing a
mixing parameter $q$
\begin{align*}
\beta_{i, C} &= \left(1 - q + q \frac{N_{i, C}}{N_i}\right) \beta \\
\beta_{i, \overline{C}} &= \beta - \beta_{i, C} \\
\lambda_{i, C} &= \beta_{i, C} \frac{I_{i, C}}{N_{i, C}}
               + \beta_{i, \overline{C}}
                 \frac{I_{i, \overline{C}}}{N_{i, \overline{C}}},
\end{align*}
where $N_{i, C}$ indicates the number of individuals of community $C$ currently
in region $i$, $N_{i, \overline{C}} = N_i - N_{i, C}$ and the other quantities
follow the same convention of notation. The intuition is as follows: When $q =
1$, everyone mixes at random inside a region just as if no countermeasure was
applied at all. At the other extreme, when $q = 0$, contacts happen only with
individuals from the same community. Intermediary values of $q$ allow us to play
with the strength of the segregation.

We evaluate the effectiveness of  {\sc DecreaseMix} for different values of the
mixing parameter $q$. Our simulations are parameterized with $\beta = 1.0$, $g =
0.5$ and $q \in \{1, 0.1, 0.01\}$; Figure \ref{fig:decmixres} shows the average
behavior of the epidemic over 10 runs. The main characteristic of this strategy
is that it delays the epidemic outbreak. However, the slopes of the two curves
at the strongest point of the epidemic are not very differentiated. As s result,
the final proportion of recovered $Q^*$ does not vary much. But by making it 10
or 100 times more likely to contact an individual of the same community, we
delay $T^*$ by approximately 5 and 16 days, respectively, on average.  Our
intuition about this phenomenon is that it takes more time for the epidemic to
reach certain communities (as they are more segregated), but once a community
sees its first case of infection, the spread is just as fast as before. We argue
that one of the main limiting factors at play here is the random mixing
assumption: if we were able to bring finer structural changes to the contact
graph, the situation would look very different.

\subsection{{\sc GoHome} strategy}

\begin{table}
  \begin{center}
  \begin{tabular}{| l | c | l |}
  \hline
  $p$ & Affected movements & Maximum   \\ \hline \hline 
  $0.1$ & $2.81  \%$ & $5.21  \%$  ($ts = 190$) \\
  $0.5$ & $15.80 \%$ & $26.12 \%$  ($ts = 316$) \\
  \hline
  \end{tabular}
  \end{center}
  \caption{Proportion of movements affected when using the {\sc GoHome} strategy for
  two different values of the compliance probability $p$. We indicate the overall
  average over the $400$ time steps, as well as the maximum value.}
  \label{tab:gohome}
\end{table}

\begin{figure}[ht]
  \centering
	\includegraphics[width=0.49\textwidth]{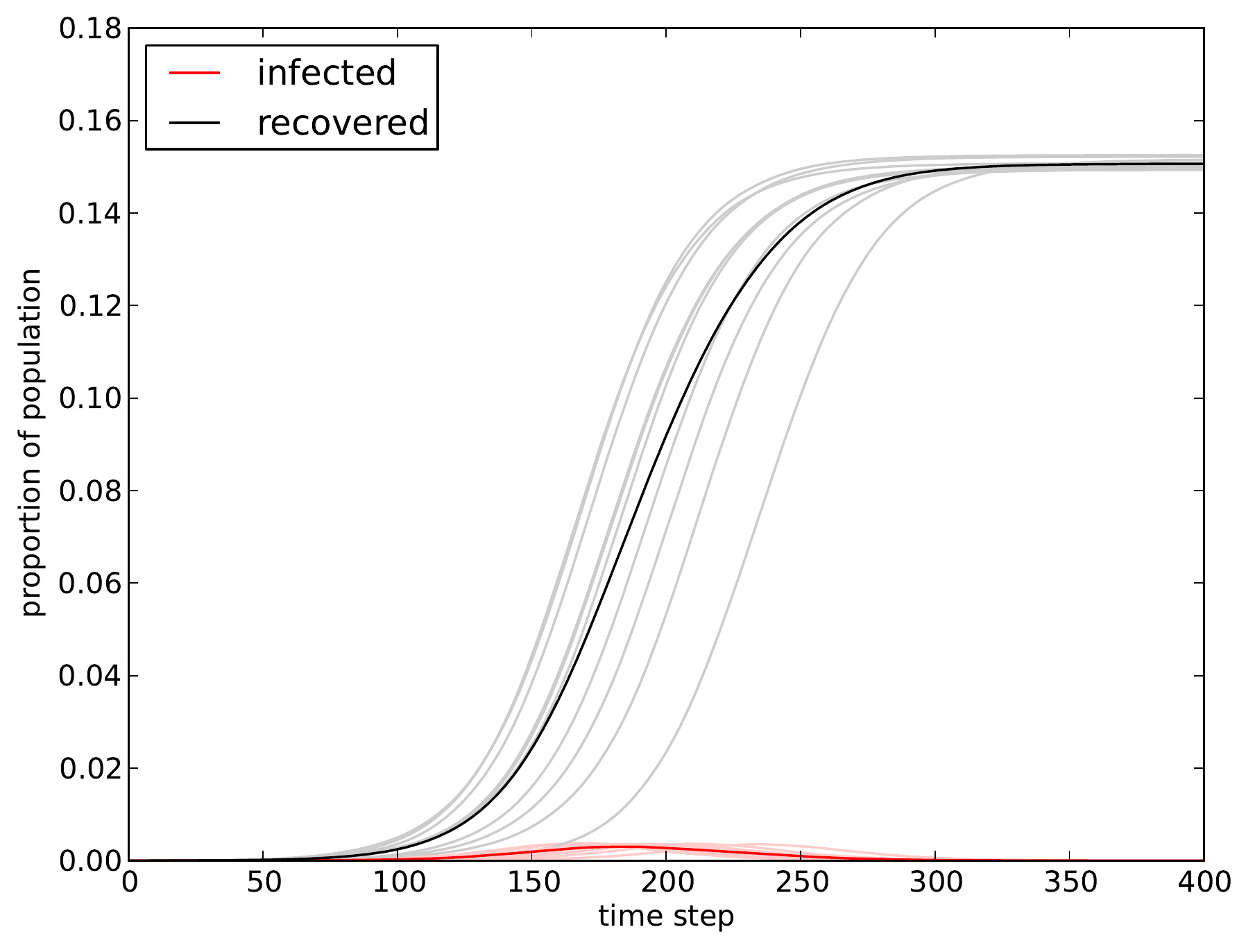}
	\includegraphics[width=0.49\textwidth]{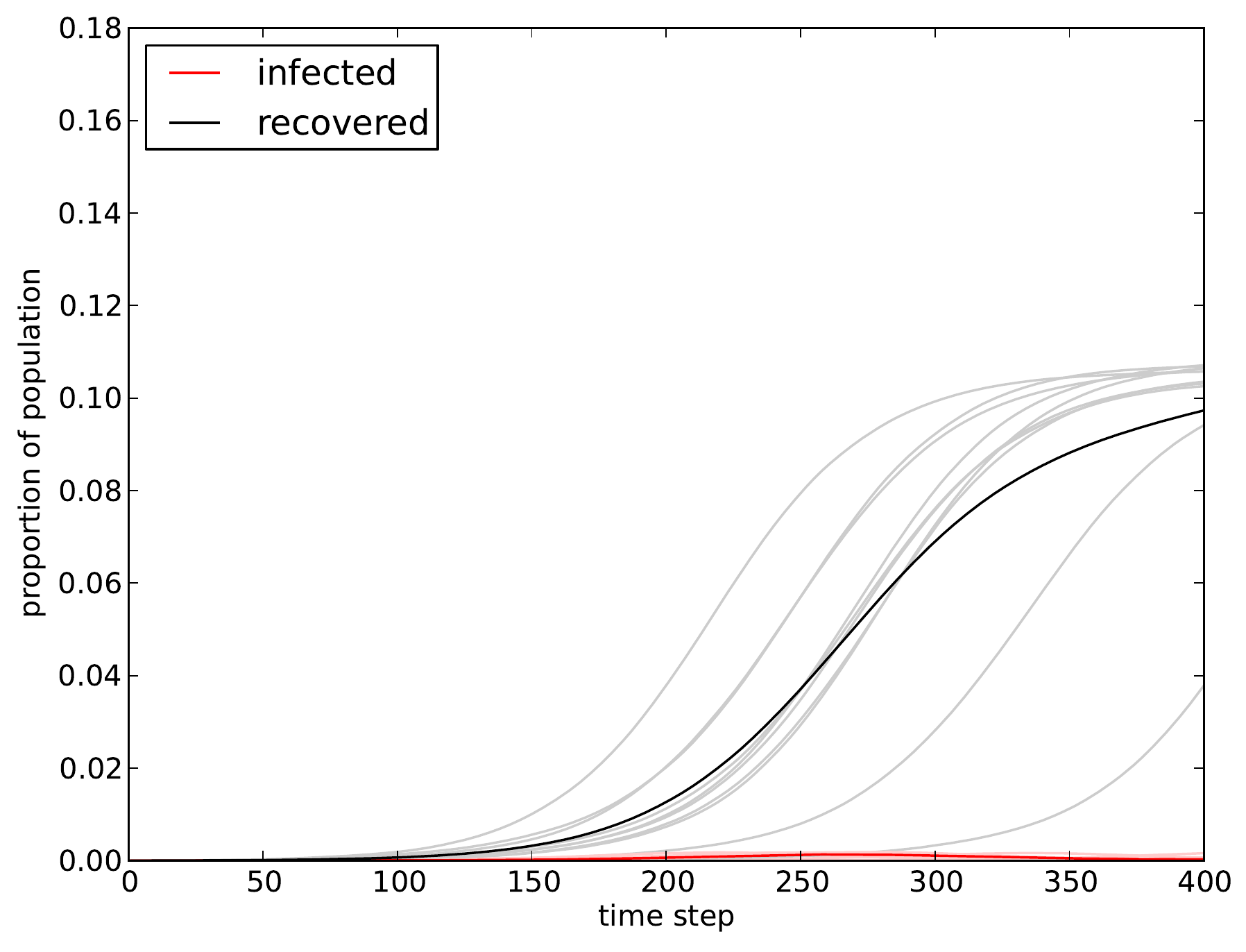}
  \caption{Shape of the epidemic under the {\sc GoHome} strategy, $\beta = 1.0$,
  $g = 0.5$. Light curves indicate individual runs, dark curves indicate
  average. On the left: $p = 0.1$, on the right: $p = 0.5$.}
	\label{fig:gohome}
\end{figure}

Our last strategy advices individuals to go home or stay home. In order to focus
the micro-measures on the most influential individuals, we assume that at each
time step, the service operator knows the proportion of susceptible, infective
and recovered individuals across locations. We suppose that before every trip,
an individual sends a request to the service that compares the proportion of
infectives in both source and destination, and recommends to go home if the
destination has a proportion of infectives, \emph{lower} than the source
location.  Individuals then comply with probability $p$.  The main intuition
behind this choice is to avoid sending infective individuals to highly
susceptible locations. Note that we keep the state-independent assumption here:
we do not know the state of the individual when sending out a recommendation.
The second important assumption is that, when the individual is at at home, the
contact probability is set to be equal to the recovery probability\footnote{When
contact and recovery probability are equal, the single-population \emph{SIR}
epidemic (under the random mixing assumption) does not develop anymore; setting
$\beta_{home} \coloneqq g$ can therefore be seen as the least change needed to
stabilize the epidemic.}, i.e.\ $\beta_{home} \coloneqq g$.  This models the
fact that there are less contacts at home, in term of accidental ones.  Mixing
is not exactly uniform anymore, and the infection probability is adapted as
follows:
\begin{align*}
\lambda_{i, loc} &= \beta_{home} \frac{I_i}{N_i} \\
\lambda_{i, vis} &= \beta \frac{I_{vis}}{N_i}
                      + \beta_{home} \frac{I_{loc}}{N_i}.
\end{align*}
Quantities with \emph{loc} and \emph{vis} subscripts correspond to individuals
whose home region is (respectively is not) $i$. Note that the contact
probability of visitors can significantly decrease in a region where the
proportion of visitors to locals is low.

This time, the effectiveness depends on the value of the compliance probability
$p$. We use again $\beta = 1.0$, $g = 0.5$ and let $p \in \{0.0, 0.1, 0.5,
0.7\}$; Figure \ref{fig:gohome} shows the behavior of the epidemic over 10 runs.
As opposed to the results obtained with the {\sc DecreaseMix} strategy, we
obtain significant improvements to $Q^*$ as $p$
increases\footnote{Unfortunately, our simulation was limited to $400$ time
steps, which is not enough to clearly show the asymptotical behavior. The claim,
however, is justified by looking at the \emph{worst} runs whose slope quickly
tends to zero.}. This observation is not surprising because by suggesting to
individuals to go home, we are directly reducing their contact probability,
which is a determining factor of the epidemic's dynamics. It is also interesting
to look at the actual number of trips that are affected (i.e.,\ cancelled)
because of the micro-measures; Table~\ref{tab:gohome} shows that a relatively
low number of trips have to be affected to noticeably impact the spread. In
summary, this strategy has the potential to be quite effective, although the
assumptions it makes deserve closer analysis.

\section{Conclusion}
\label{sec:conclusion}

In this paper, we explore the novel idea of using mobile technology in order to
mitigate the spread of human-mediated infectious diseases.  We motivate the
concept of \emph{mobile micro-measures} that consist of personalized behavioral
recommendations given to individuals.  By affecting, even partially, individual
behaviors, we are able to globally impact the epidemic propagation. These mobile
micro-measures have several original properties; they are adaptive, target
individuals at the microscopic level and provide a rich set of mitigation
methods.
Using the data provided for the Orange D4D challenge \cite{blondel2012data}, we
first develop a realistic mobility model for the population of Ivory Coast.
Then, we incorporate it into an epidemic model based on \emph{SIR} in order to
simulate the epidemic propagation, while taking into account population
mobility. Taking advantage of this framework, we propose and evaluate three
concrete strategies used to generate micro-measures.  Our strategies weaken the
epidemic's intensity, successfully delay its peak and, in one case,
significantly lower the total number of infected individuals.

These preliminary results allow us to identify several research avenues. First,
random mixing is the most limiting assumption. Being able to change the
structure of human contacts at a finer level is a key component of more advanced
micro-measures. The mobile call graph is an example of a source of information
about social contacts, one that is readily available to mobile phone operators.
Second, beyond our preliminary strategies, it is highly important to deepen our
understanding of the key ingredients that make mobile micro-measures effective
yet minimally restrictive.  In parallel to mobile micro-measures, the
availability of large-scale mobility data opens up new research directions in
epidemiology: a more precise characterization of the relation between epidemic
spread and human mobility patterns is an interesting topic we would also like to
investigate in the future.

To conclude, we firmly believe that data-driven and personalized measures which
take advantage of mobile technology are an important step towards effective
epidemic mitigation.

\section*{Acknowledgements}

We would like to thank Vincent Etter for his insightful comments and feedback
about this paper.

\bibliography{report}

\end{document}